\documentclass[%
 reprint,
 amsmath,amssymb,
 aps,
]{revtex4-1}

\usepackage{graphicx}
\usepackage{dcolumn}
\usepackage{bm}
\usepackage{color}

\begin{document}


\title{Repeated sequential learning increases memory capacity via effective decorrelation in a recurrent neural network}

\author{Tomoki Kurikawa}
 \email{kurikawt@hirakata.kmu.ac.jp}
 \affiliation{Kansai medical university, Shinmachi 2-5-1, Hirakata, Osaka, Japan}

\author{Omri Barak}
\affiliation{Rappaport Faculty of Medicine and Network Biology Research Laboratories, Technion - Israeli Institute of Technology, Haifa, Israel}

\author{Kunihiko Kaneko}%
\affiliation{%
 Graduate school of arts and sciences, University of Tokyo, Komaba 3-8-1, Meguro-ku, Tokyo, Japan
}%

\date{\today}

\begin{abstract} 
Memories in neural system are shaped through the interplay of neural and learning dynamics under external inputs. 
By introducing  a simple local learning rule to a neural network,  we found that the memory capacity is drastically increased by sequentially repeating the learning steps of input-output mappings.
The origin of this enhancement is attributed to the generation of a Psuedo-inverse correlation in the connectivity. This is 
associated with the emergence of spontaneous activity that intermittently exhibits neural patterns corresponding to  embedded memories.  
Stablization of memories is achieved  by a distinct bifurcation from the spontaneous activity  under the application of  each input.
\end{abstract}

\maketitle

Through sequential learning, the brain learns to appropriately respond to various inputs.
In neural system, synaptic connections are modified to shape neural dynamics such that the applied stimulus and desired response are adequately represented therein. After learning, the stimulus is represented according to the shaped neural dynamics\cite{Blumenfeld2006,Bernacchia2007,McKenzie2013,Dunsmoor2015,Driscoll2017}.
How memories are successively embedded into neural dynamics through the interplay between the neural dynamics and learning process is a crucial question in neuroscience.

To understand the representation of memories in neural system, associative memory models are often studied.
In conventional models\cite{Amari1977,Hopfield1984,Amit1987}, multiple memories are designed to be embedded into corresponding attractors and are generated by a simple learning rule. 
In spite of their success, however, the interplay between neural dynamics and learning has not been taken into account: 
when learning a new memory, the change in a connection was determined only by the memory pattern, independently of the already-shaped neural dynamics.  

In contrast, we previously proposed a novel associative memory model\cite{Kurikawa2012,Kurikawa2013} that incorporates interactions between neural dynamics and the learning process. 
In these studies, however, each pattern was only presented once during learning, and existing memories are gradually eroded as new patterns are learned.

In the present Letter, we first introduce a theoretical formulation for a sequential and repeated learning process which interacts with neural dynamics. 
By studying this learning process, we investigated if all memories are able to be successfully stored by repeated learning.
If so, we then address what kind of neural-network structure enables such enforcement and how memories are represented in neural dynamics upon input. We also study spontaneous dynamics without input which is suggested to involve computation in neural system\cite{Orban2016,Berkes2011,Buesing2011,Ma2006,Litwin-Kumar2012,Hennequin2018}.

We consider a model that consists of $N$ continuous rate-coding neurons to memorize $M$ Input-Output (I/O) mappings.
The activity $\bm{x}=\{x_i\}$ $(i=1,2,\cdots, N)$ is set between -1 and 1 and evolves according to
\begin{equation}
  \dot{x_{i}} =\tanh{(\beta(\sum_{j \neq i}^{N} J_{ij}x_{j} + \gamma\eta^{\mu}_{i}))} - x_{i}, \label{eq:neuro-dyn}
\end{equation}
where $J_{ij}$ denotes a connection from the $j$-th to $i$-th neuron, an $N$ dimensional vector $\bm{\eta}^{\mu}$ is an input pattern, $\gamma$ being its strength, and $\mu$ is the index of I/O mappings to be learned. 
In the following section, $\gamma$ and $\beta$ are set at 1.0 and 4.0, respectively, unless otherwise stated.

For each input $\bm{\eta}^{\mu}$, we set an $N$ dimensional vector $\bm{\xi}^{\mu}$ as a target.
These input and target patterns are generated as random $N$-bit binary patterns, with probabilities $P(\xi_{i}=\pm 1)=P(\eta_{i}=\pm 1)=1/2$.
In the presence of each input $\bm{\eta}^{\mu}$, the corresponding target $\bm{\xi}^{\mu}$ is required to be recalled, i.e.,
an attractor matching $\bm{\xi}^{\mu}$ is generated. 
The learning process is required to modify the connectivity  $J_{ij}$ to enable the network recall the targets.

Previously\cite{Kurikawa2013}, we showed such a memory structure is formed through a simple learning rule. 
To make repeated sequential learning, we added a decay term
\begin{equation}
  \dot{J_{ij}} = (\epsilon/N) (\xi^{\mu}_{i} - x_{i})(x_{j} - h_{i}J_{ij}), \label{eq:syn-dyn-LRNN}
\end{equation}
where $h_{i}=\sum_{j \neq i} J_{ij}x_{j}$.
We use a learning rate $\epsilon=0.03$ unless otherwise stated.
According to this learning rule, $d (\sum_{j \neq i} J_{ij}^{2})/dt \propto (1-\sum_{j \neq i} J_{ij}^{2})$.
Thus, this rule preserves $\sum_{j \neq i} J_{ij}^{2}=1$, if initially $\sum_{j \neq i} J_{ij}^{2}=1$:
$J_{ij}$ takes a binary value with probabilities $P(J_{ij}=\pm (N-1)^{-1/2})=1/2$ before learning.
Diagonals of $\bm{J}$ are set at zero during the whole process.
The learning process stops automatically when the neural activity matches the target, 
because $\dot{J_{ij}}=0$; otherwise, the learning process continues. 
Here we imposed $M$ I/O maps successively: 
an input is applied to learn a target and after learning the map was completed, another input is applied to learn another target.
The learning process for each single I/O map is called a learning step and denoted $T$.
During the learning process, maps are applied in the order for $T \leq M$ steps.
Then, they are randomly applied for $T>M$ steps.

\begin{figure}[t]
  \begin{center}
     \includegraphics[width=80mm]{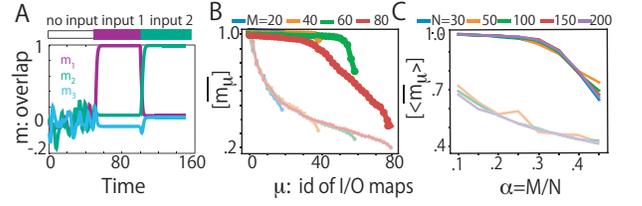}
    	\caption{
    		Recall performance.
		A: $m_{\mu} = \Sigma_{i} x_{i} \xi_{i}^{\mu} /N$ ($\mu={1,2,3}$) are plotted.  
		An input 1 and then 2 is applied for $50 \leqq t<100$ and $100 \leqq t <150$, respectively. 
		B: $[\overline{m_{\mu}}]$ averaged over time, networks and trials are plotted against the index of maps after learning $M$ maps.
		After sorting overlap $\overline{m_{\mu}}$  ($\mu=1,,,M$) by its value, we averaged sorted overlaps over 5 trials for each of 5 networks with $N=200$.
		C: memory performance for different $N$ is plotted against $M$ normalized by $N$. 
		We averaged $m_{\mu}$ over all $\mu$ of 5 trials for each of 5 networks.
		In B an C, shadowed lines represent overlaps after $T=M$ learning steps, while bold lines are those after $T=30M$ steps.
	}
  \label{fig:rcl-perf}
  \end{center}
\end{figure}

Fig. \ref{fig:rcl-perf}A shows the recall processes in response to two input patterns after learning.
$\bm{\eta}^{1}$ is applied from $t=50$ to $100$. Under the input, neural dynamics are modified and the neural state converges to $\bm{\xi}^{1}$.
When $\bm{\eta}^{2}$  is applied instead of the input 1, the neural dynamics are modified differently and, then, the neural state converges to $\bm{\xi}^{2}$.
A neural state that provides a desired target pattern is an attractor.
Here, these two I/O maps are successfully recalled.

We first analyzed how repeated learning enhances the memory capacity.
For this purpose, we computed the temporal average of an overlap of neural activity $\bm{x}$ with a target, 
$[\overline{m_{\mu}}] = \Sigma_{i} \overline{x_{i}} \xi_{i}^{\mu} /N$ in the presence of input $\bm{\eta}^{\mu}$ $(\mu=1,,,M)$. 
$\overline{...}$ and $[...]$ represent the temporal average and average over network and trials in Fig. \ref{fig:rcl-perf}B.
At this stage, networks can recall only one or two targets perfectly and overlaps with other targets decrease rapidly, independently of $M$.
After learning these targets more and more times ($T=30M$), however, recall performance increases and targets of about 60 are recalled perfectly.
Networks fail to memorize target patterns beyond $M=60$.
Thus, $M=60$ indicates the limit of memory. 

To evaluate this memory capacity in detail, we calculated the averaged overlap $<[m_{\mu}]>_{\mu}$ over maps, networks and trials, and plotted it for different $N$ in Fig \ref{fig:rcl-perf}C,
where $<...>_{\mu}$ represents average over index of I/O maps.
After $T=M$ learning steps, the average overlap decreases rapidly, while, after $T=30M$ learning steps, the overlap is maintained at around unity up to $\alpha=M/N=0.3$.
Therefore, the capacity of the present model is estimated to be $\alpha_{c}=0.3-0.35$.
To explore dependence of the memory on $T$, we examined $[<m_{\mu}>_{\mu}]$ for different $T$. 
We found that the memory capacity increases monotonically as $T$ increases and saturates around $T=20M$ (See Fig. S1).
Thus, we studied behavior for $T=M$ and $30M$ as typical samples in the earlier and later stage of learning.

Enhancement in the memory capacity after iterative learning is not trivial, but depends on the learning speed $\epsilon$ and $\beta$. 
As shown in Fig. S1, the memory capacity is decreased as $\epsilon$ increases and $\beta$ decreases.
Especially, the memory capacity for $\epsilon \geq 1$ and $\beta \leq 1$ is almost one, which cannot be increased by repeated learning.
This result indicates the need for an adequate relation between the timescales of neural and learning dynamics,  as well as the nature of neural dynamics,  is important to enhance memory capacity through repeated learning.

\begin{figure}[t]
  \begin{center}
     \includegraphics[width=80mm]{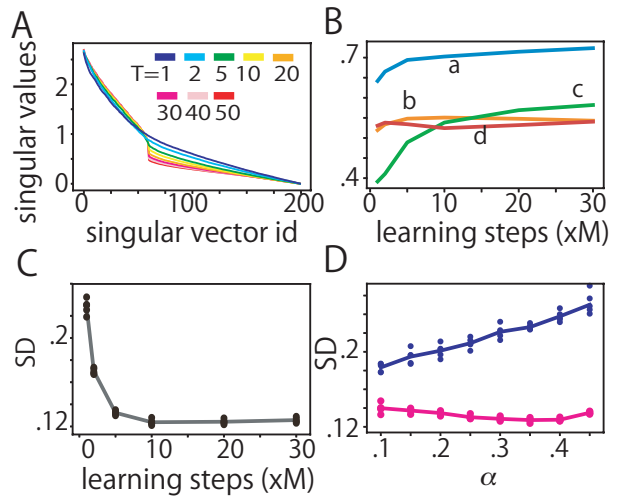}
    \caption{
      A: Singular value distribution of 5 learned networks for $N=200$ and $M=60$ by singular value decomposition for different learning steps T. 
      The other panels are for the same networks.
      B: $<\Sigma_{\mu} (s^{\kappa,\mu})^{2}>_{0 \leq \kappa \leq M-1}$ for $s=a,b,c,d$ are plotted during learning process.
	C: Standard deviation (SD) of non-diagonal elements of $S$ during learning process. Different points indicate different networks.
	D: SD of  $\bm{\xi}^{\mu} \bm{J} \bm{\xi}^{\nu} /N$ is plotted for $T=M$ in blue and for $T=30M$ in magenta.
      }
  \label{fig:diag-data}
  \end{center}
\end{figure}

Next, we examined the nature of $\bm{J}$ shaped throughout the learning process and its relevance to the enhancement in the memory capacity.
To this end, we calculated singular values of the connectivity for different learning steps.
A learned connectivity $\bm{J}$ is decomposed as $\bm{J}=\bm{U \Sigma V^{t}}$.
$\bm{V}^{t}$ is a transpose matrix of $\bm{V}$ with $\Sigma$ as diagonal matrix whose elements are singular values.
The values are plotted in the order of their magnitude for $N=200, M=60$ in (Fig \ref{fig:diag-data}A).
They decrease continuously for earlier learning steps, while,  after long learning,  there appears a large discontinuity at 60.
For different $N$, the singular value always show a discontinuous drop at $M$ at the later learning stage.
This means $M$ left and right singular vectors become dominant in the connection throughout learning\footnote{The remaining $N-M$ singular vectors still enhance the memory performance. Actually, a matrix consisting of only $M$ dominant vectors shows a small decrease in the performance as shown in Fig S2. }.

Recalling the connections in the Hopfield model\cite{Hopfield1984} ($J=\Sigma_{\mu=0}^{M-1} \bm{\xi}^{\mu} (\bm{\xi}^{\mu})^{t}/N$) and 
our previous models \cite{Kurikawa2012,Kurikawa2013} ($J=\Sigma_{\mu} (\bm{\xi}^{\mu}-\bm{\eta}^{\mu})(\bm{\xi}^{\mu}+\bm{\eta}^{\mu})^{t}/N$), 
we hypothesize that these $M$ vectors consist mainly of linear combinations of $\bm{\xi}$ and $\bm{\eta}$ and the other $N-M$ left and right singular vectors are in the normal space to these combinations.
To examine this hypothesis, we used $a^{\kappa, \mu}=\Sigma_{i} u_{i}^{\kappa}\xi_{i}^{\mu}/N, b^{\kappa, \mu}=\Sigma_{i} u_{i}^{\kappa}\eta_{i}^{\mu}/N, c^{\kappa, \mu}=\Sigma_{i} v_{i}^{\kappa}\xi_{i}^{\mu}/N,$ and $d^{\kappa, \mu}=\Sigma_{i} v_{i}^{\kappa}\eta_{i}^{\mu}/N$. 
Here, $u_{i}^{\kappa}$ and $v_{i}^{\kappa}$ are $i$-th elements of $\kappa$-th left and right singular vectors, respectively.
Contributions of $\bm{\xi^{\mu}}$ and $\bm{\eta^{\mu}}$ to $\bm{u}^{\kappa}$ ($\bm{v}^{\kappa}$) are roughly estimated by $a^{\kappa, \mu}$ and $b^{\kappa, \mu}$ ($c^{\kappa, \mu}$, and $d^{\kappa, \mu}$), respectively
\footnote{``roughly'' means $O(1)$ and we discarded terms of $O(N^{-1/2})$, since targets and inputs are not exact normal orthogonal basis ($\Sigma_{i} \xi_{i}^{\mu} \eta_{i}^{\nu} = O(N^{-1/2})$) }.
We measured $<\Sigma_{\mu} (a^{\kappa,\mu})^{2}>_{0 \leq \kappa \leq M-1}$, the average contribution of targets to one of dominant $M$ left singular vectors and 
also the corresponding quantities for $b,c$ and $d$ in Fig.\ref{fig:diag-data}B.
All of the values are much higher than chance level $M/N=0.3$ meaning that the dominant $M$ vectors mainly consist of targets and inputs.
Particularly $a^{\kappa, \mu}$ and $c^{\kappa, \mu}$ increases with learning.

We also found that $a^{\kappa,\mu}$ is highly correlated with $b^{\kappa,\mu}$, while $c^{\kappa,\mu}$  is correlated to $d^{\kappa,\mu}$ (Fig. S2).
Thus, dominant $M$ left and right singular vectors are decomposed as
\begin{equation}
	\bm{u}^{\kappa} \sim \Sigma_{\mu} a^{\kappa,\mu}(\bm{\xi}^{\mu}+k^{\mu}\bm{\eta}^{\mu}), \bm{v}^{\kappa} \sim \Sigma_{\mu} c^{\kappa,\mu}(\bm{\xi}^{\mu}+l^{\mu}\bm{\eta}^{\mu}),
\end{equation}
where $k^{\mu}$ ($l^{\mu}$) is the correlation coefficient between $a^{\kappa,\mu}$ and $b^{\kappa,\mu}$, ($c^{\kappa,\mu}$ and $d^{\kappa,\mu}$).
We found also that $a^{\kappa,\mu}$ is highly correlated with $c^{\kappa,\mu}$ across $\kappa$ for a given $\mu$, but not with $c^{\kappa,\nu}$ (Fig. S2).
By this analysis, in total, $\bm{J}$ is decomposed as
\begin{equation}
	\bm{J} \sim \Sigma_{\mu} S_{\mu \nu} (\bm{\xi}+k^{\mu} \bm{\eta}^{\mu})(\bm{\xi}^{\nu}+l^{\nu} \bm{\eta}^{\nu})^{t},
\end{equation}
where $S_{\mu \nu} = \Sigma_{\kappa} \sigma_{\kappa} a^{\kappa \mu} c^{\kappa \nu}$ and $\sigma^{\kappa}$ is an $\kappa$-th singular value.
Note that, to enhance recall performance, non-diagonal terms $S_{\mu \nu}$ ($\mu \neq \nu$) should be small.
In our model, actually, they are much smaller than the diagonal ones of $S$, since there is no correlation between $a^{\kappa,\mu}$ and $c^{\kappa,\nu}$.
Additionally, these non-diagonal terms are further reduced as learning progresses, as shown in Fig. \ref{fig:diag-data}C. 

To achieve optimal memory capacity, it is generally believed that inverse matrix of correlation between targets has to be introduced into the connectivity \cite{Personnaz1986,Kanter1987,Diederich1987} to reduce the interference due to correlation between targets.
In the case of our model, $\bm{J} \bm{\xi}^{\lambda} = S_{\lambda \lambda}(\bm{\xi}^{\lambda}+k^{\lambda}\bm{\eta}^{\lambda}) + O_{\lambda}$, 
where the interference term $O_{\lambda}$ is defined as $\Sigma_{\mu, \nu} S_{\mu \nu}(\bm{\xi}^{\mu}+k^{\mu}\bm{\eta}^{\mu})(\bm{\xi}^{\nu}+l^{\nu}\bm{\eta}^{\nu})^{t} \bm{\xi}^{\lambda}$ ($\nu \neq \lambda$ and $\mu \neq \lambda$).

Instead of obtaining the exact form of the connectivity\footnote{For it, detailed analysis to the order of  $O(N^{-1/2})$ is necessary, which is not easy due to the same order of noise generated by correlation between any two random patterns.}, we focus on whether the learned connectivity effectively decorrelates the patterns.
Recalling that, in the standard Hopﬁeld network corresponding to the case that $S$ is a diagonal matrix, the standard deviation of $\bm{\xi}^{\mu} \bm{J} \bm{\xi}^{\nu} /N$ ($\mu \neq \nu$) follows $O((\alpha/N)^{1/2})$, whereas it follows $O(N^{-1/2})$ for the Pseudo-inverse correlation matrix \cite{Personnaz1986,Kanter1987,Diederich1987}.
We, hence, measured $\bm{\xi}^{\mu} \bm{J} \bm{\xi}^{\nu} /N$ ($\mu \neq \nu$) and estimated its dependence on $N$ and $\alpha$ in Fig. \ref{fig:diag-data}D and in Fig. S2 for the present connection matrix shaped by learning.
We found that the standard deviation at the earlier stage of learning (at $T=M$), it follows $O((\alpha/N)^{1/2})$, 
but at the later stage of learning it turns to follow $O(N^{-1/2})$ (for $T=30M$).
This result implies that our learning rule effectively shapes the inverse correlation matrix into the connectvity throughout the learning process to optimally reduce interference.

\begin{figure}[t]
  \begin{center}
     \includegraphics[width=80mm]{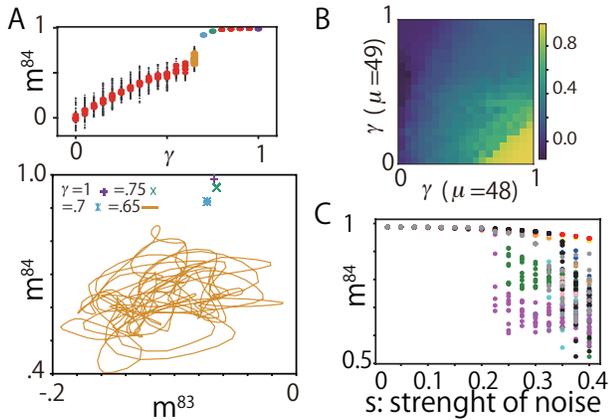}
    \caption{
    	A: In the upper panel, the bifurcation diagram of overlap with $\bm{\xi^{84}}$ through input strength for $N=300, T=30M$ and $\alpha=0.3$.
	The overlaps averaged over time for 10 trials are plotted as 10 red points. 
	Small black dots represent snapshots of the overlaps over time.
	In the lower panel, attractors for different strength are displayed. Neural activity is projected onto 2-dimensional plane by taking its overlaps.
	Different color lines represent attractors for different strength $\gamma$ indicated above.
	B: Phase diagram of $\overline{m_{48}}$ for different input patterns ($\eta^{48}$,$\eta^{49}$)is plotted.
	C: bifurcation diagram against noise strength is plotted. Different color indicates different realization of random perturbed pattern.
	}
  \label{fig:bif_analys}
  \end{center}
\end{figure}

Next, we analyzed how memories are represented after the inverse correlation matrix is shaped.
We first focus on modification of neural dynamics against input strength $\gamma$.
In Fig. \ref{fig:bif_analys}A, we plot a bifurcation diagram against $\gamma$ for $T=30M$.
Neural activity for $\gamma=0$, i.e., spontaneous neural activity, oscillates around the origin. 
As $\gamma$ increases, it moves towards a target while maintaining the oscillation amplitude.
At a certain strength, the attractor of neural dynamics bifurcates from oscillation to a fixed point corresponding to the target.
Neural dynamics projected onto 2-D plane is plotted around the bifurcation point in Fig. \ref{fig:bif_analys}.
Neural activity with a large-amplitude oscillation reduces into a fixed point corresponding to the target between $\gamma=0.65$ and $0.7$.
Beyond the bifurcation point, the fixed point stays around the target as  $\gamma$ is increased.
Thus, neural activity corresponding to target recall is clearly distinguished from other activities through a bifurcation and is stable against change in $\gamma$ beyond the bifurcation point.

We then explored the behavior of neural activity against the mixture of two learned inputs.
As an example, the phase diagram of $\overline{m_{48}}$ against strength of two learned inputs ($\mu=48,49$) is shown in Fig. \ref{fig:bif_analys}B.
The fixed point corresponding to $\bm{\xi}^{48}$ shapes a distinctive phase, at boundary of which bifurcation from the target fixed point to oscillating dynamics occurs (Fig. S3).
The fixed point of the other target ($\mu=49$) provides similar phase diagram (Fig. S3). 
As an input pattern is changed to increase $\gamma$ in the form of $\gamma \bm{\eta}^{48} + (1-\gamma) \bm{\eta}^{49}$ ($0 \leqq \gamma \leqq 1$), 
the attractor bifurcates from the fixed point of $\bm{\xi}^{49}$ to oscillating neural activity close to $\bm{\xi}^{49}$ and to that close to $\bm{\xi}^{48}$, then, to the fixed point of $\bm{\xi}^{48}$.
These results show that a target is represented as a distinctive phase of the fixed point which is separated by the bifurcations from the attractor with the oscillation.

We asked how robust these memories are against perturbation in inputs.
To examine the robustness, we applied quenched random noise with strength $s$ to the original input patterns,  
as $\bm{\eta}'^{\mu} = \bm{\eta}^{\mu} + s\bm{\zeta}$ ($\bm{\zeta}$ is an $N$-dimensional vector whose elements are random number from uniform probability distribution $[-1,1]$) and analyzed stability of the neural activity that recalls the target (Fig. \ref{fig:bif_analys}C).
For small $s$, the fixed point of the target is insensitive to the noise and remains around unity.
Beyond the bifurcation point, the fixed point is collapsed into neural activity showing oscillation.

\begin{figure}[t]
  \begin{center}
     \includegraphics[width=80mm]{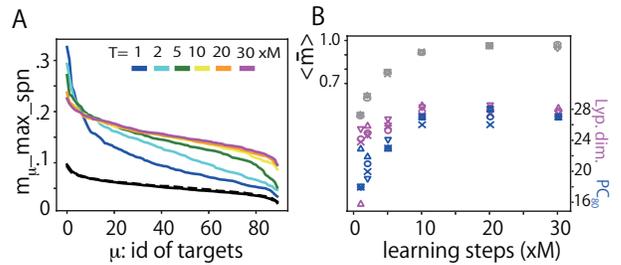}
    \caption{
        Modification of the spontaneous activity for four networks, indicated by different marks, through learning for $M=90$ and $N=300$. A: maximum overlap averaged over different I/O maps $<\text{max}_{0<t<1000}m_{\mu}(t)>_{\mu}$ is plotted. Black fill and dotted lines indicate overlaps with input and random patterns for reference. B:recall performance and Lyapunov dimension of spontaneous activity are plotted in gray and magenta, respectively. index of PC component at which normalized cumulative summation of PC-contribution is over 0.8 ($PC_{80}$) in blue.
     }
  \label{fig:spn}
  \end{center}
\end{figure}

To close the analysis of neural dynamics, we explored how spontaneous activity is related to the recall performance through learning.
For earlier learning step, spontaneous activity shows chaotic dynamics that intermittently approach and depart from targets in Fig. S4A.
Here, only a few targets each of which is successfully recalled upon input are approached (as well as their opposite patterns due to parity symmetry in our model) in Fig. \ref{fig:spn}A and Fig. S4B.
For later step, the spontaneous activity also approaches targets, but, here, many targets are more equally approached.
We further analyzed neural dynamics by using principal components (PC) analysis and measuring Lyapunov dimension in Fig. \ref{fig:spn}B.
We found that the variation of the spontaneous activity is larger and more chaotic, when learning progresses and recall performance is improved.
Thus, the spontaneous activity is constrained on lower dimension along axis connecting target patterns for lower recall performance, 
whereas that is distributed more isotropically against target patterns across higher dimensions for higher performance.

To confirm this relation between spontaneous activity and recall performance generally,
we examined the spontaneous activity for different $\epsilon$ in Fig. S4.
For smaller $\epsilon$, recall performance is higher and the spontaneous activity shows high-dimensional distribution which is close to all of the targets.
As $\epsilon$ increases, in contrast, recall performance decreases and the spontaneous activity turns to be low-dimensional, approaching to only a few targets which are perfectly recalled. Finally, for quite large $\epsilon$ (=5), the spontaneous activity turns to be a few of fixed points which are target patterns and only these target are successfully retrieved.
These results support the relation between the spontaneous activity approaching the targets and recall performance.

To sum, by studying neural networks that memorize I/O maps, we have shown how repeated learning stabilizes each memorized state and enhances memory capacity via the interplay between neural dynamics and learning. 
In usual sequential learning, e.g., gradient descent method\cite{Rumelhart1985,Williams1989} and palimpsest memory\cite{Amit1994,Brunel1998,Fusi2007}, connections are slowly shaped. The network's output moves in the direction of the desired target, but does not match it after a single step. In contrast, in the present study, connections are modified such that the network generates the correct target after each step in one shot. Thus, we can analyze how targets are embedded in neural dynamics and how the representation of these targets changes through learning. 
Interaction between neural dynamics and learning was investigated to reveal how neural representation is shaped in several studies \cite{Bernacchia2007,Bernacchia2014,Blumenfeld2006,Siri2008,Galtier2011,Kim2008}. These studies, however, did not focus on parametric effects of neural dynamics (e.g., the gain parameter) and learning (e.g., the learning speed) on learning performance and representation of memories. 

Spontaneous activity which intermittently reproduces  stimulus-evoked patterns is commonly reported in visual\cite{Kenet2003,Berkes2011} and auditory\cite{Luczak2009} cortices.
Theoretical studies\cite{Zenke2015,Hartmann2015,Miconi2016,Litwin-Kumar2014,Bernacchia2014} demonstrated how the spontaneous activity is shaped through learning. Our study provides another simple learning rule to form such spontaneous activity.
Further, we showed a relation between features of spontaneous activity and recall performance - consistent with its interpretation as a prior distribution in terms of probabilistic inference\cite{Orban2016,Berkes2011,Buesing2011,Ma2006}.  
More generally, properties of neural dynamics relevant for information processing were investigated\cite{Toyoizumi2011,Bertschinger2004,Legenstein2007,Sussillo2009}, and the edge of chaos was suggested as an appropriate regime.
Our model suggests that high-dimensional chaos with intermittent visits to learned patterns is suitable to produce appropriate targets in response to inputs.
The role of such itinerant dynamics\cite{Kaneko2003} has been discussed over decades\cite{Tsuda1992,Skarda1987,Rabinovich2008}, 
and the present study clearly demonstrates it.
 
The Pseudo inverse model\cite{Personnaz1986,Kanter1987,Diederich1987} can achieve higher memory capacity $1.0N$ than the standard Hopfield network $0.14N$ \cite{Hopfield1984,Amit1987}.
In this model, the inverse correlation matrix of memories is included in the connectivity to reduce interference among memories in recall, and non-local information is required to shape this connection.
Further, Diedrich\cite{Diederich1987} proved the local learning rule 
$\Delta J_{ij} = (1/N)(\xi_{i}^{\mu}-\Sigma_{j} J_{ij} \xi_{j}^{\mu})\xi_{j}^{\mu}$ 
can shape such a connectivity after repeated learning. 
In our model, if we focus only on the relaxation dynamics in the vicinity of $\bm \xi^{\mu}$, a fixed point of neural dynamics in eq. \ref{eq:neuro-dyn}, 
which is given by $x_{i} \sim \tanh{(\beta(\sum_{j \neq i}^{N} J_{ij}\xi_{j}+\eta^{\mu}_{i}))}$.
Then, the learning rule in eq. \ref{eq:syn-dyn-LRNN} takes a similar form with the Diedrich rule, by neglecting the decay term.
This may partially explain why our local, repeated learning shapes the connection matrix to include the inverse correlation matrix and enhances the memory capacity.

In the present study, in contrast to the ordinary associative memory\cite{Hopfield1984,Amit1987,Amit1994,Brunel1998,Fusi2007}, each memory is recalled through an input-induced bifurcation from the spontaneous neural activity.  After repeated learning, the spontaneous activity and the fixed point of the recalled memory state are distinguished discontinuously through this bifurcation, resulting in the stability of memory against a perturbed input pattern. 
Although modulation of neural dynamics by input is analyzed in some studies\cite{Minai1998,Rajan2010,Rubin2015}, our study suggests that the memory state is represented as a robust and distinct phase against the parameter space of input strength.
This discrete representation of memory often observed in auditory\cite{Bathellier2012}, olfactory\cite{Niessing2010} cortices and Hippocampus\cite{Wills2005}. In these cortices, neural activity patterns are discretely switched between two memory states depending on the intensity of sensory inputs and/or ratio of mixture of two different inputs. Our model provides a simple learning rule to form such memory representations and gives a prediction in the terms of spontaneous activity properties and memory performance.

\section{Aknowledgement}
We thank David Colliaux for fruitful discussion.
This work was partly support by KAKENHI (no. 18K15343) and Hitachi The university of Tokyo for funding.

\section*{Supplemental materials}

\subsection{Change in recall performance during learning process}
We studied how memory performance is changed through learning.
We plotted $[\overline{m^{\mu}}]$ against $\mu$ with the increase of $T$.
It is sorted in the order of magnitude of the overlap in Fig S1A(i).
For early learning stage, only a few targets are stored, while, for the later stage, the number of targets perfectly recalled increases rapidly.
Here, we measured $[<\overline{m_{\mu}}>_{\mu}]$ as recall performance in A(ii).
For $N=200$ and $M=60$, we plot the recall performance against leaning step $T$. 
The capacity increases rapidly up to $T=10M$ and almost saturates at $T=20M$.

\subsection{recall performance for different $\epsilon$ and $\beta$}
For $\epsilon = 0.03$ and $\beta=4$, the memory capacity is enhanced through repeated learning.
We explored its dependence on different parameters, especially, $\epsilon$ and $\beta$.
$\epsilon$ is a time scale of learning process relative to that in neural dynamics.
We plotted capacity curve for various $\epsilon$ in Fig.S1B.
As $\epsilon$ increases, the number of patterns which are successfully recalled decreases and 
for $\epsilon > 1$, only one pattern is recalled.
We also explored dependence of the recall performance on $\beta$.
Generally, in randomly coupled neural network models, attractors change from fixed points to chaos with the increase in $\beta$.
We plotted the recall performance for different $\beta$ in Fig. S1C.
As $\beta$ increases, the recall performance is increased.
For $\beta < 1$, only one or two memories are recalled successfully.
These results show that relationship between timescales of neural dynamics and learning process is crucial to shape successful memories.

\subsection{relationship between $a^{\kappa, \mu},b^{\kappa, \mu},c^{\kappa, \mu}$ and $d^{\kappa, \mu}$}
Scatter plot of $(a^{\kappa,\mu},b^{\kappa,\mu})$ is displayed in Fig S2A(i).
$a^{\kappa, \mu}$ is negatively correlated with $b^{\kappa,\mu}$.
We also show a scatter plot of ($a^{\kappa, \mu}, c^{\kappa, \mu}$) in Fig S2A(ii).
$a^{\kappa, \mu}$ is positively correlated with $c^{\kappa,\mu}$.
In the Right panel in Fig S2A, we plot ($a^{\kappa, \mu}, c^{\kappa,\nu}$) ($\mu \neq \nu$).
There is no correlation between them.

\subsection{dependence of $\bm{\xi} \bm{J} \bm{\xi}^{t} $ on $N$}
We plot the standard deviation (SD) of $\Sigma_{i,j} \xi_{i}^{\mu} J_{ij} \xi_{j}^{\nu} /N$ for different $N$ in Fig S2B.
SD both in earlier and later stages of learning are scaled as $N^{-1/2}$.

\subsection{Representation of targets}
We explored the behaviors of the overlap with the target against inputs.
In Fig S3A, we show the bifurcation of the overlap with the target 48 against strength of the input 48 under the presence of input 49 in addition to Fig \ref{fig:bif_analys}B.
the bifurcation of the overlap with the target 49 is plotted in Fig S3C, whereas the bifurcate diagram for inputs 48 and 49 are shown in Fig S3B. 
All the results support that recall of the target pattern is represented as a distinctive phase of the corresponding fixed-point attractor and as separated from oscillating neural activity.

\subsection{Spontaneous activity}
We analyzed how the nature of spontaneous activity is changed through learning.
Spontaneous activity shows chaotic behavior intermittently approaching some targets in Fig. S4A.
For earlier learning, we found clear correlation between recall performance $<\overline{m}>$ and maximum overlap $ <\text{max}_{0<t<1000} m_{\mu} (t)>_{\mu}$ as shown in Fig. S4B.
A few targets which show nearly perfect recall performance are closely approached by the spontaneous activity.
For later learning, in contrast, there appears no clear correlation. Almost all targets show perfect recall performance and their closeness (the maximum overlap) is distributed in middle value.

Next, we explored spontaneous activity for different $\epsilon$.
As $epsilon$ decreases and recall performance increases (Fig. S1B and Fig. S4E), the spontaneous activity is distributed broader (Fig. S4D and Fig. S4F) and more chaotic (Fig. S4F).
This relation between the spontaneous activity and recall performance is consistent with that for different learning step.
For quite larger $\epsilon$,  some fixed points, instead of chaotic dynamics, are shaped, one of which corresponds to the latest trained network in Fig. S4G.

\begin{figure*}[t]
	\centering
     \includegraphics[width=130mm]{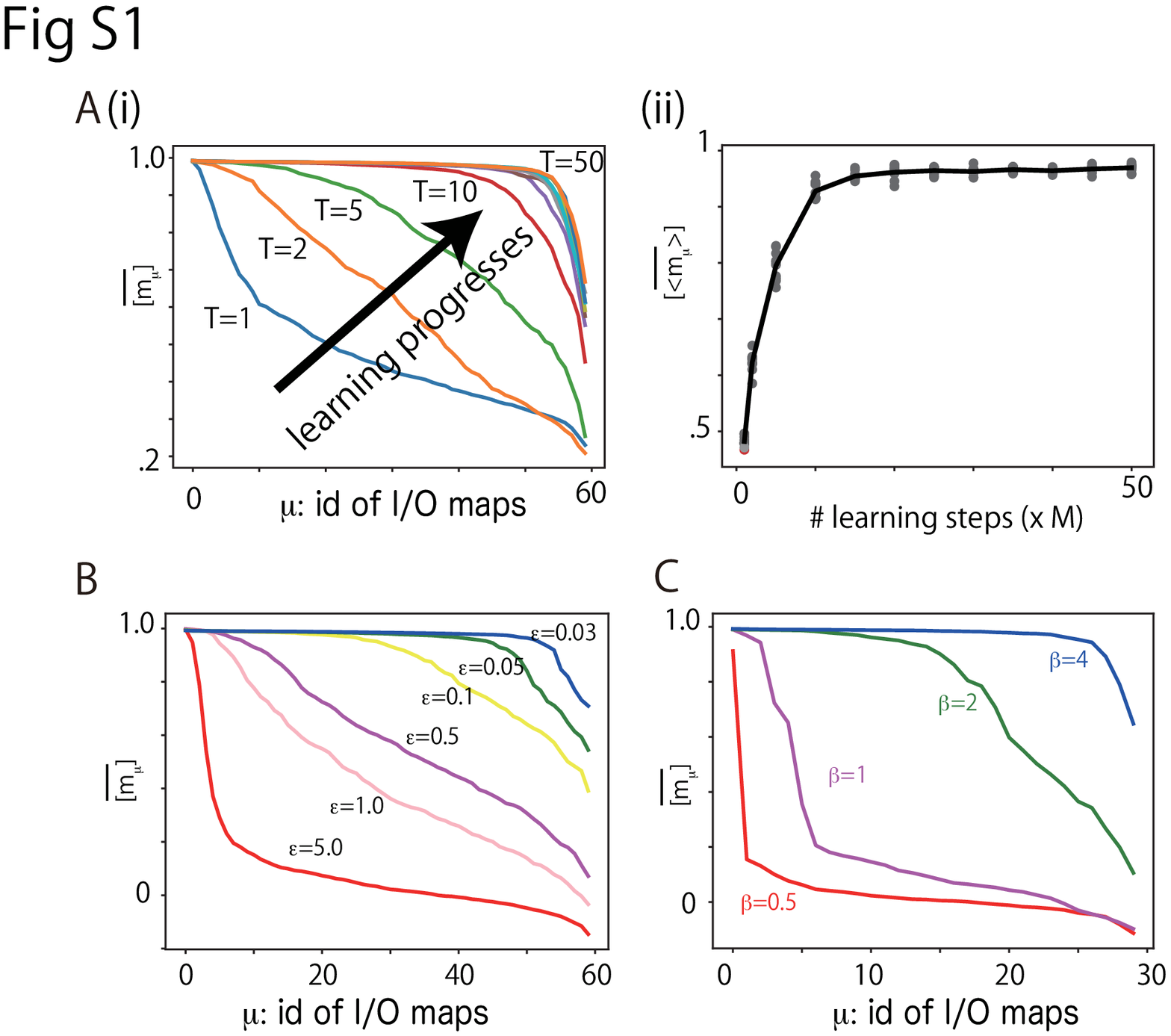}
	\newline
	\begin{flushleft}
    Figure S1:  A(i): $[\overline{m_{\mu}}]$ as a function of $\mu$ is plotted for different $T$. 
    $[\overline{m_{\mu}}_{\mu}]$ averaged over time, networks and trials are plotted for $N=200$ and for $\alpha=0.3$. 
    In (ii), $[<\overline{m_{\mu}}>_{\mu}]$ by averaging over $\mu$ as a function of $T$ is plotted .
    B: $[\overline{m_{\mu}}]$ is plotted for $T=30M$ and for different $\epsilon$.
    C: $[\overline{m_{\mu}}]$ is plotted for $T=30M$,$N=100$, $\alpha=0.3$ and for different $\beta$.
       
\end{flushleft}
\end{figure*}

\begin{figure*}[t]
	\centering
     \includegraphics[width=130mm]{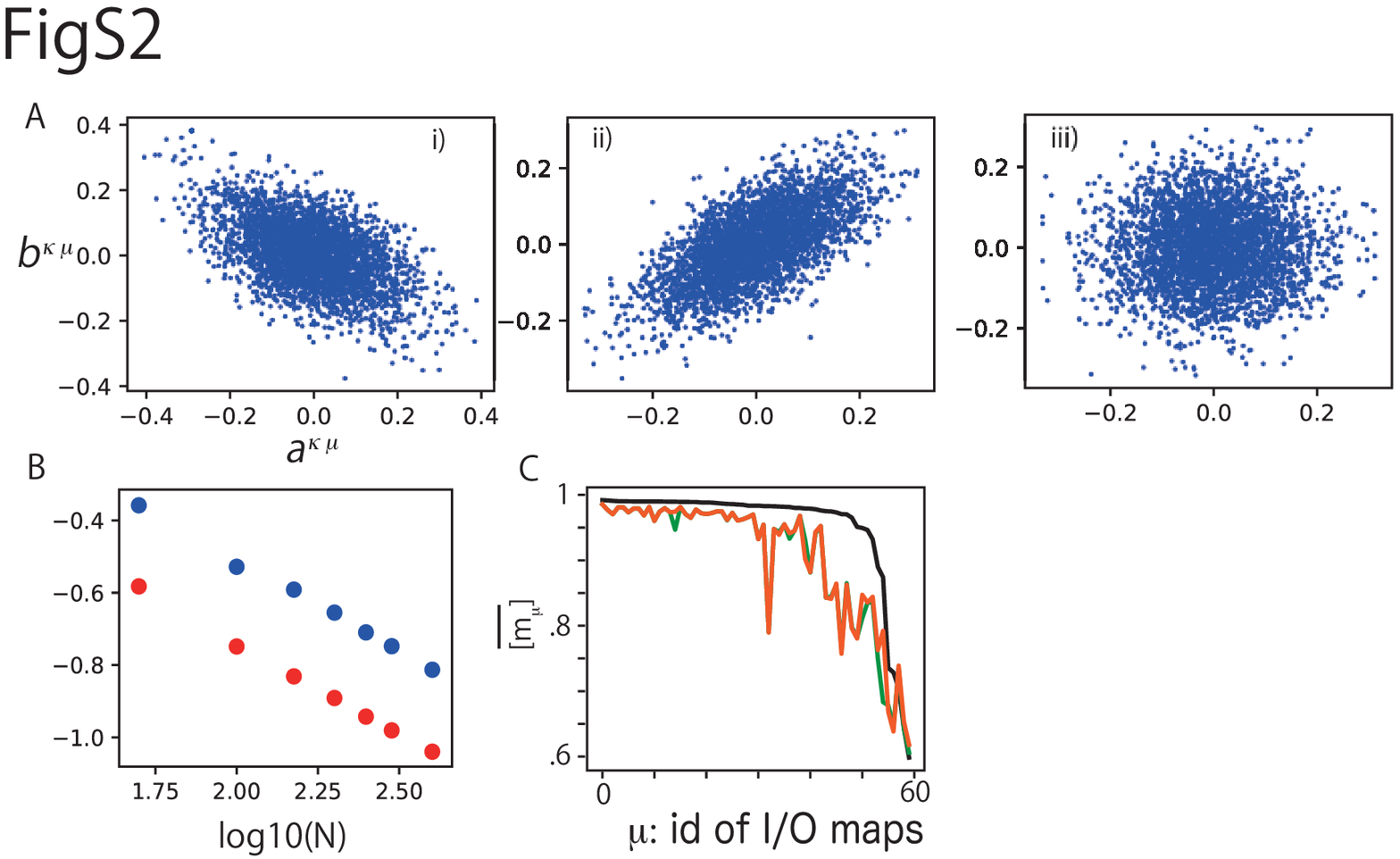}
	\newline
	\begin{flushleft}
	Figure S2 \newline
	A: Left panel, scatter plot between $a_{\kappa, \mu}$ and $b_{\kappa,\mu}$ for $0\leq \kappa, \mu \leq M-1$ and $M=60, T=30M$.
	Center panel, scatter plot between  $a_{\kappa, \mu}$ and $c_{\kappa,\mu}$ for the same parameters as Left panel.
	Right panel, scatter plot between $a_{\kappa, \mu}$ and $c_{\kappa,\nu}$ ($\mu \neq \nu$) for the same  parameters as Left. We plot points for randomly selected 60 pairs of $(\mu,\nu)$.
	B: The standard deviation of $\Sigma_{i,j} \xi_{i}^{\mu} J_{ij} \xi_{j}^{\nu} /N$ for different $N$. Blue dots are from $T=M$ and red ones are from $T=30M$.
	C: Recall performance for the matrix consisting of only $M$ dominant singular vectors. Green line indicates the performance for this matrix, while red one indicates the performance for the matrix which is re-scaled so that $\sum_{j \neq i} J_{ij}^2 = 1$. Black line represents the performance for the original matrix for reference.
\end{flushleft}
\end{figure*}

\begin{figure*}[t]
	\centering
     \includegraphics[width=130mm]{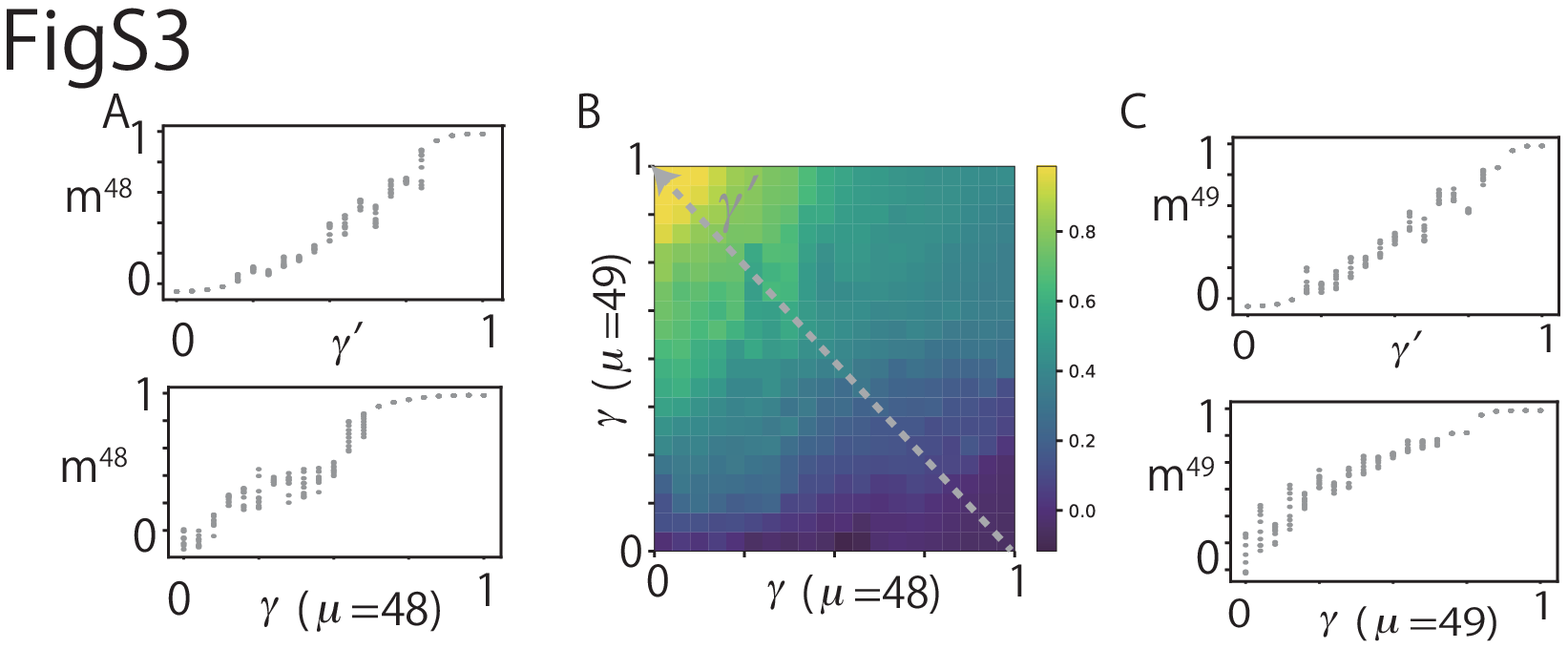}
	\newline
	\begin{flushleft}
	Figure S3\newline
	A:Bifurcation of the overlap with the target 48. In the upper panel, the bifurcation against $\gamma\bm{\eta^{48}}+(1-\gamma)\bm{\eta^{49}}$ is plotted. 
	In the lower panel, the bifurcation against $\gamma\bm{\eta^{48}}$ is plotted.
	B: Phase diagram of the overlap with the target 49. In the upper panel, the bifurcation against $\gamma\bm{\eta^{49}}+(1-\gamma)\bm{\eta^{48}}$ is plotted. 
	In the lower panel, the bifurcation against $\gamma\bm{\eta^{49}}$ is plotted.
\end{flushleft}
\end{figure*}

\begin{figure*}[t]
	\centering
     \includegraphics[width=130mm]{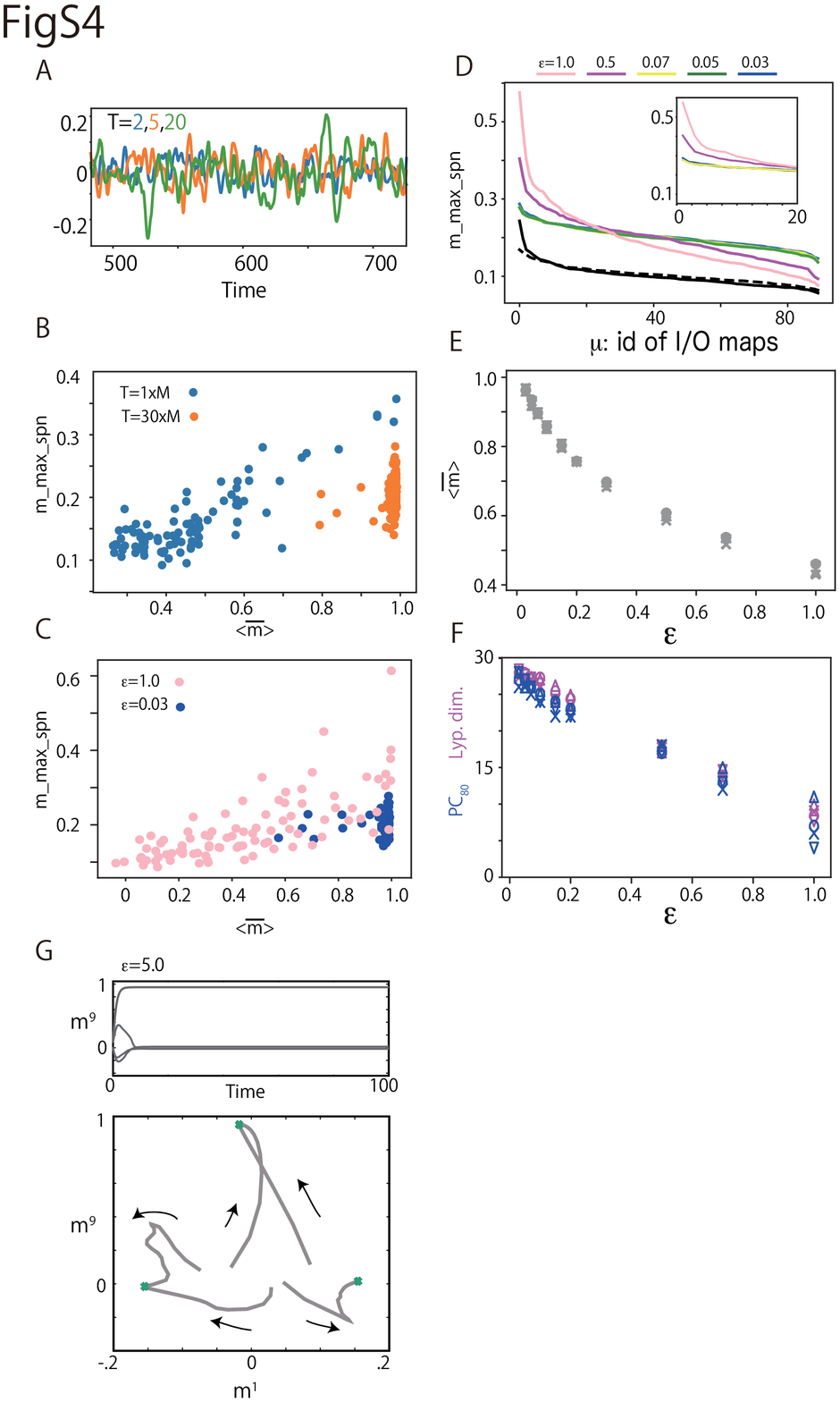}
     \newline
     Caption is in the next page
\end{figure*}
\addtocounter{figure}{-1}
\begin{figure*} [t!]
    
	\begin{flushleft}
	
	Figure S4 \newline
	    
        A: Overlaps of spontaneous activities without input $m_{45}$ are plotted from t=500 to 700 for $T=2M,5M$ and $T=20M$ in blue, orange, and green, respectively, on the left panel. Here, $N=300$ and $M=90$ as well as in the following panels. On the right panel, probability density functions (p.d.f) of these overlaps in longer interval ($0<t<1000$) and their standard deviations ($pm \sigma$)are also plotted as bars. 
        B: Scatter plot of maximum overlap of the spontaneous activity with a target against its recall performance. Dots in blue and orange are for $T=M$ and $T=30M$, respectively.
        C: Scatter plot same as in B. Dots in blue and pink are for $\epsilon=0.03$ and $\epsilon=1.0$, respectively.
        D: Maximum overlaps of the spontaneous activity with targets are plotted for different $\epsilon$. Black filled and dotted lines indicate overlap with input and random patterns, respectively, for reference.
        E and F: Recall performance in E and Lyapnov dimension and $PC_{80}$ in F are plotted.
        G: Spontaneous activity for $\epsilon=5.0$. 
    Upper panels: 5 time series of the overlap with target 9 from 5 random initial points, which is the latest trained pattern.
    Lower panels: the same neural dynamics with upper panels projected into 2-dimensional space.  
\end{flushleft}
\end{figure*}


\end{document}